\documentclass[%
 reprint,
superscriptaddress,
 amsmath,amssymb,
 aps,
]{revtex4-1}

\usepackage{graphicx}
\usepackage{dcolumn}
\usepackage{bm}
\usepackage{color}

\begin{document}


\title{A platform for electrically pumped \\
polariton simulators and topological lasers}

\author{Holger Suchomel}
\email{holger.suchomel@physik.uni-wuerzburg.de}
\affiliation{Technische Physik, Wilhelm-Conrad-R\"ontgen-Research Center for Complex
Material Systems, Universit\"at W\"urzburg, Am Hubland, D-97074 W\"urzburg,
Germany}
\author{Sebastian Klembt}
\email{sebastian.klembt@physik.uni-wuerzburg.de\\ H. Suchomel and S. Klembt contributed equally to this work.}
\affiliation{Technische Physik, Wilhelm-Conrad-R\"ontgen-Research Center for Complex
Material Systems, Universit\"at W\"urzburg, Am Hubland, D-97074 W\"urzburg,
Germany}
\author{Tristan H. Harder}
\affiliation{Technische Physik, Wilhelm-Conrad-R\"ontgen-Research Center for Complex
Material Systems, Universit\"at W\"urzburg, Am Hubland, D-97074 W\"urzburg,
Germany}
\author{Martin Klaas}
\affiliation{Technische Physik, Wilhelm-Conrad-R\"ontgen-Research Center for Complex
Material Systems, Universit\"at W\"urzburg, Am Hubland, D-97074 W\"urzburg,
Germany}
\author{Oleg A. Egorov}
\affiliation{Technische Physik, Wilhelm-Conrad-R\"ontgen-Research Center for Complex
Material Systems, Universit\"at W\"urzburg, Am Hubland, D-97074 W\"urzburg,
Germany}
\author{Karol Winkler}
\affiliation{Technische Physik, Wilhelm-Conrad-R\"ontgen-Research Center for Complex
Material Systems, Universit\"at W\"urzburg, Am Hubland, D-97074 W\"urzburg,
Germany}
\author{Monika Emmerling}
\affiliation{Technische Physik, Wilhelm-Conrad-R\"ontgen-Research Center for Complex
Material Systems, Universit\"at W\"urzburg, Am Hubland, D-97074 W\"urzburg,
Germany}
\author{Ronny Thomale}
\affiliation{Institut f\"ur Theoretische Physik, Universit\"at W\"urzburg,
  Am Hubland, D-97074 W\"urzburg, Germany}
\author{Sven H\"ofling}
\affiliation{Technische Physik, Wilhelm-Conrad-R\"ontgen-Research Center for Complex
Material Systems, Universit\"at W\"urzburg, Am Hubland, D-97074 W\"urzburg,
Germany}
\affiliation{SUPA, School of Physics and Astronomy, University of St Andrews, St Andrews
KY16 9SS, United Kingdom}
\author{Christian Schneider}
\affiliation{Technische Physik, Wilhelm-Conrad-R\"ontgen-Research Center for Complex
Material Systems, Universit\"at W\"urzburg, Am Hubland, D-97074 W\"urzburg,
Germany}

\begin{abstract}

Two-dimensional electronic materials such as graphene and transition metal dichalgenides feature unique electrical and optical properties due to the conspirative effect of band structure, orbital coupling, and crystal symmetry. Synthetic matter, as accomplished by artificial lattice arrangements of cold atoms, molecules, electron patterning, and optical cavities, has emerged to provide manifold intriguing frameworks to likewise realize such scenarios. Exciton-polaritons have recently been added to the list of promising candidates for the emulation of system Hamiltonians on a semiconductor platform, offering versatile tools to engineer the potential landscape and to access the non-linear electro-optical regime. In this work, we introduce an electronically driven square and honeycomb lattice of exciton-polaritons, paving the way towards real world devices based on polariton lattices for on-chip applications. Our platform exhibits laser-like emission from high-symmetry points under direct current injection, hinting at the prospect of electrically driven polariton lasers with possibly topologically non-trivial properties.

\end{abstract}

\pacs{Valid PACS appear here}
\maketitle


\section{\label{sec:level1}INTRODUCTION}

\begin{figure*}
  \includegraphics[width=0.99\textwidth]{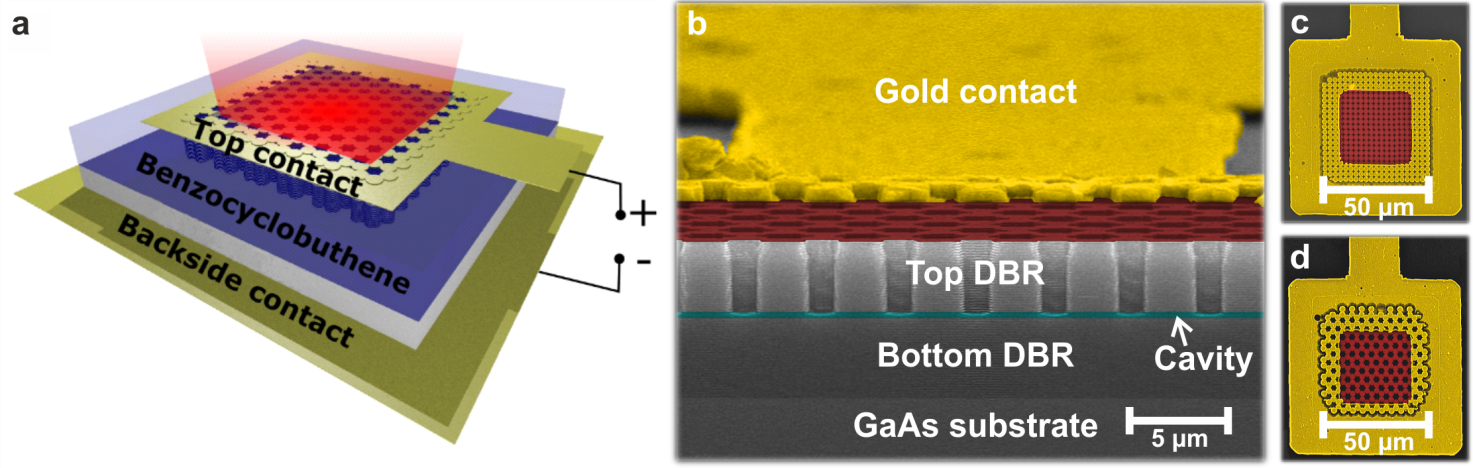}
  \caption{(a) Schematic  of the investigated device showing the half-etched polariton lattice, planarized with benzocyclobutene with top and bottom electrical contacts. (b) Scanning electron microscope (SEM) image of the processed device with a view tilted above the cleaved edge, cutting the device in half. A gold contact (yellow) has been deposited on an etched two-dimensional lattice (red) in order to inject a current into the microcavity structure. The quantum wells (QWs) have been placed in the field maximum inside a $\lambda$-cavity (cyan). The n-contact is deposited on the plane backside of the GaAs substrate. To avoid etching damage to the QWs, the etching depth has been adjusted in order to stop close above the cavity region. (c-d) SEM images of the investigated lattice devices in top view configuration. The deposited gold contact (yellow) surrounds the etched lattices (red) with an overlap of one to two lattice constants. The whole lattice field has an edge length of $\sim$ 50 $\mu$m with (c) depicting the investigated square lattice and (d)  the honeycomb lattice.}
\end{figure*}

Microcavity exciton-polaritons (polaritons) are composite bosonic particles originating from the strong coupling between excitons and microcavity photons \cite{Weisbuch1992}. Their bosonic nature, strong nonlinearities, as well as their convenient accessibility by angular-resolved photo- or electroluminescence (PL and EL) spectroscopy put them in the focus of contemporary fundamental research devoted to macroscopic quantum phases in solid state systems, culminating in the observation of collective bosonic behaviour manifest in Bose-Einstein condensation \cite{Imamoglu1996,Kasprzak2006,Balili2007} and superfluidity \cite{Amo2009}. Furthermore, even the classical operational limit of polariton lattices has triggered significant interest due to the prospect of enabling detailed model engineering towards artificial topological matter \cite{Karzig2015, Bardyn2015, Nalitov2015}, which has recently peaked in the experimental implementation of optically driven topological polariton lasers \cite{Jean2018} and polariton Chern insulators \cite{Klembt2018}.

Previously studied in a variety of material systems \cite{Deng2002,Christopoulos2007,Lai2012,Cohen2010} and microcavity architectures \cite{Bajoni2007,Gessler2014,Zhang2015,Dietrich2017}, the currently highest quality devices for polaritons are based on epitaxially grown III-V distributed Bragg reflectors (DBRs) with embedded GaAs quantum wells (QWs) \cite{Sun2017}, where techniques derived from semiconductor nanotechnology could be fruitfully employed \cite{Amo2016,Berloff2017,Ohadi2009,Gao2015,Schneider2017,Jacqmin2014}. In order to accomplish a tunable lattice arrangement of microcavities, micropillar etching is employed to implement deep potentials along with a careful tuning of the pillar proximity and overlap in order to enable lattice hybridization (hopping) \cite{Bayer1999,Dasbach2003}. It has led to the design of polariton band structures on the square \cite{Kim2011,Winkler2015}, Lieb \cite{Klembt2017,Whittaker2018}, Kagome \cite{Masumoto2012}, and honeycomb lattice \cite{Jacqmin2014,Kusudo2013}. Along with other degrees of freedom constituting candidates for synthetic matter such as microwave cavities, photonic crystals, waveguides, and electric circuits, polariton lattices are evolving into a promising platform for emulating topological states of matter \cite{Karzig2015, Bardyn2015, Nalitov2015}. As a distinguishing advantage towards opto-electronic applications, however, polaritons allow not only optical but also electrical driving, which is likely indispensable for any vision of on-chip technology, and likewise, real world applications. Techniques for direct current injection into polariton states \cite{Tsintzos2008}, including the nonlinear regime of bosonic condensation \cite{Schneider2013,Bhattacharya2013} have been developed, and represent an important tool for practical “plug-and-play” implementations \cite{Tsotsis2009,Suchomel2017}.

In this work, we present a technology platform facilitating electrical injection into arbitrary two-dimensional lattice geometries hosting polaritons, and study the band structure formation in an electrically pumped square- and honeycomb lattice in the strong coupling limit, both in the linear and non-linear regime. We demonstrate that we can electrically drive $\sim$ 250 lattice sites in the tight-binding (TB) limit, marking a step change in engineering of photonic lattices in general. We observe polariton band structures that exhibit good agreement with the TB description of the lowest energy band (S-band) for a square and honeycomb lattice, the latter exhibiting its distinct $K$ and $K'$ points, and the corresponding Dirac cone dispersion reminiscent of graphene \cite{Neto2009}.
\section{\label{sec:level2}EXPERIMENTAL SETUP}
\subsection{Implementation}

The polariton device is based on a pin-doped, vertically emitting microcavity structure with 27 Al$_{0.20}$Ga$_{0.80}$As/AlAs mirror pairs in the p-doped top distributed Bragg reflector (DBR), 31 Al$_{0.20}$Ga$_{0.80}$As/AlAs mirror pairs in the n-doped bottom DBR, and a single stack of four GaAs QWs embedded in the field antinode of the intrinsic Al$_{0.40}$Ga$_{0.60}$As $\lambda$-wavelength-thick cavity. The cavity structure was grown by molecular beam epitaxy on a n-doped GaAs substrate. We characterized the wafer via white-light reflectance measurements at 10 K by using the spatial variation of the cavity resonance and extracted a Rabi splitting of $\hbar \Omega_R$ = $(5.4 \pm 0.1)$\,meV in the presence of the built-in potential from the unbiased pin-junction. The quality factor of the cavity was extracted from PL measurements on a far-red-detuned area of the wafer, yielding a value of Q $\sim$ 15,000.
A detailed description of the sample growth, doping profile and basic optical properties can be found in the Supplementary Material.\\
\indent The electrically injected polaritonic lattices, schematically depicted in Fig. 1a, were implemented on the microcavity structure by a series of processing steps. First, an AuGe-Ni-Au alloy was evaporated and annealed on the backside of the substrate serving as the n-contact. In order to generate the 2D polariton lattices, square as well as honeycomb lattices were defined by electron beam lithography. After the deposition of BaF$_2$ as an etching mask and the subsequent lift-off step, the structures were etched using electron cyclotron resonance-reactive ion etching. Here, the etching depth is calibrated such that only the top DBR is etched and the cavity with the embedded QWs remains intact (see scanning electron microscope (SEM) image Fig. 1b, Supplementary Material, Fig. S1c). Current injection is provided by fabricating electrically contacted frames overlapping with one to two lattice constants of the lattice. In order to generate a platform for the contacting step at the p-doped side of the structure, the sample was planarized with the transparent polymer benzocyclobutene (BCB) using a thin layer of sputtered Si$_3$N$_4$ as an adhesive layer (Supplementary Material, Fig. S1d). Besides serving as a platform for the contact level, the BCB is also used as an electrical insulation between different devices and to prevent oxidation of the exposed lattice sidewalls. Finally, the contact level on top of the planarized sample was defined via optical lithography and evaporation of an Au-Cr alloy. The following lift-off step completes the device processing. Side, as well as top view images of the contacted square and honeycomb lattices are shown in Figs. 1b-d. The width of a full lattice array amounts to 50 $\mu$m. 

\subsection{Experiment}
We characterize our processed devices by momentum resolved PL and EL spectroscopy. For the PL measurements, we excite our sample by a 2 ps pulsed Ti:sapphire laser with a repetition rate of 82 MHz. For each device the laser energy was tuned to the reflectance minimum of the first high-energy Bragg mode around $\sim$ 84 meV above the emission energy. In the case of the EL measurements, the sample was excited by applying a DC bias between the backside and top contact using a standard voltage source. The injected current was determined by measuring the voltage drop across a series resistance of 1 k$\Omega$. In both excitation configurations, the luminescence is collected using a Fourier spectroscopy setup with a Czerny-Turner monochromator and a Peltier-cooled 1024x1024 px CCD operating at -100 $^{\circ}$C. By motorized scanning of the last imaging lens we can collect the full dispersion information in the $k_x$- and $k_y$-directions. All following experiments on the processed devices have been carried out at a sample temperature of $\sim$ 4 K.
\section{\label{sec:level2}RESULTS AND DISCUSSION}
\subsection{Honeycomb Lattice}

\begin{figure}
  \includegraphics[width=0.99\linewidth]{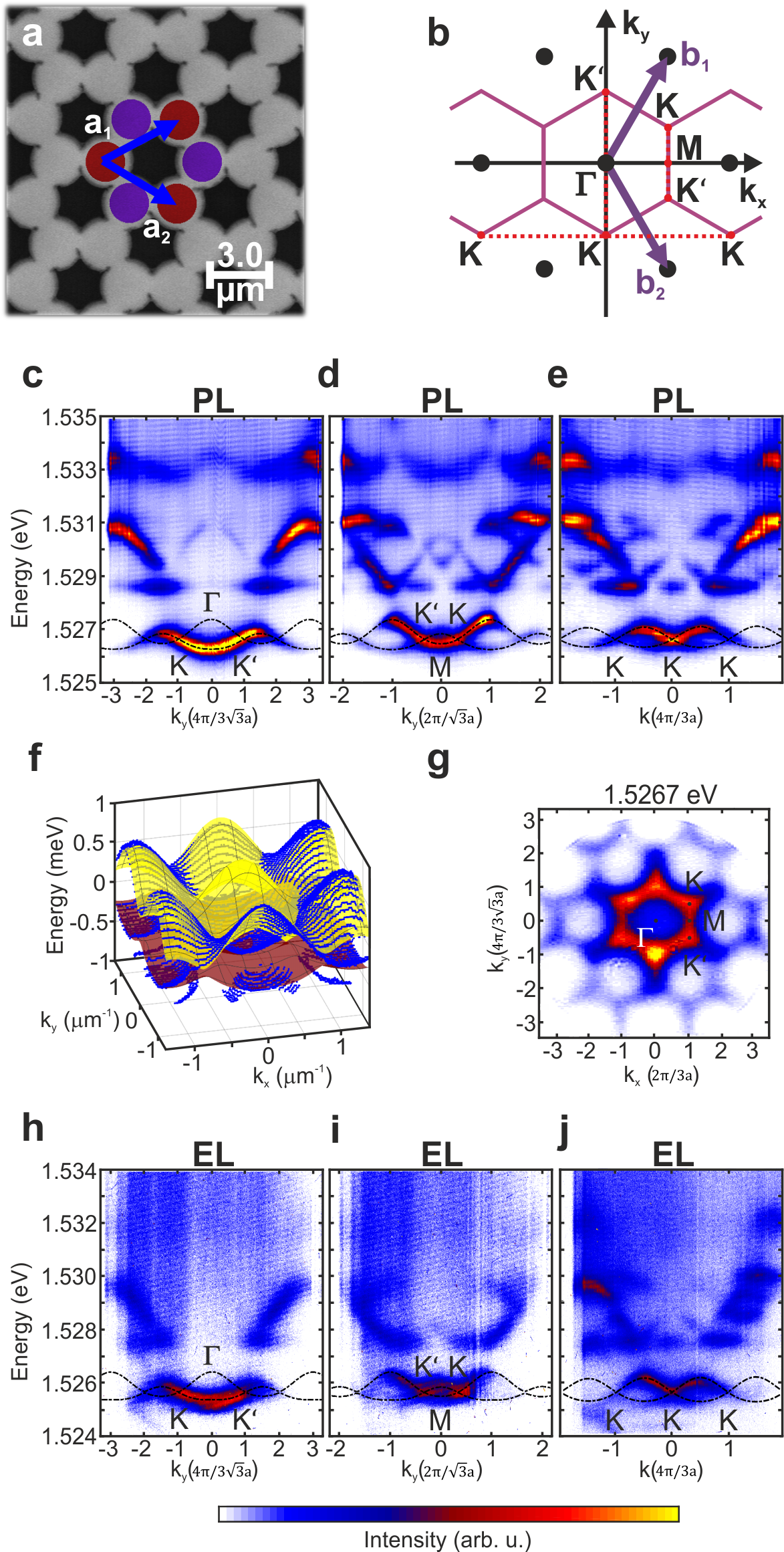}
  \caption{
  (a) Scanning electron microscope image of the investigated honeycomb lattice and the corresponding reciprocal space (b). (c-e) PL dispersion in the K-$\Gamma$-K$^{\prime}$ (c), K$^{\prime}$-M-K (d) and K-K-K direction (e) of the 1st Brillouin zone (BZ), showing the distinct Dirac cone dispersion around the K-points. The dashed line is a tight-binding (TB) calculation reproducing the S-band of the dispersions. The excitation density amounts to $\sim$ 10 Wcm$^{-2}$. (f) Hyperspectral imaging of the 1st BZ with the peak positions fitted to the dispersion spectra (blue) and the TB simulation of the lower (red) and upper S-band (yellow). (g) Energy cut in reciprocal space at the center of the S-band, showing a clear honeycomb geometry with the K and K$^{\prime}$ points. (h-j) EL emitted by the very same honeycomb lattice in the K-$\Gamma$-K$^{\prime}$ (h), K$^{\prime}$-M-K (i) and in the K-K-K (j) direction of the 1st BZ. The TB calculation uses the same fit parameter as for the PL measurement.}
\end{figure}

A lattice geometry which attracted a substantial interest in emulating two-dimensional Hamiltonians is the graphene lattice, made up of single sites arranged in a honeycomb configuration. The SEM image in Fig. 2a depicts the honeycomb lattice studied in this article with the two-element base as well as the primitive lattice vectors being highlighted. The corresponding reciprocal lattice is shown in Fig. 2b. The 1st Brillouin zone (BZ) includes the high symmetry points M, K, and K$^{\prime}$ at the edge and $\Gamma$ in the center. The ratio between the pillar diameters of $d=3.0$ $\mu$m to the lattice constant of $a=2.7$ $\mu$m amounts to $\sim$ 0.9. We characterize the optical properties of the lattice by non-resonant laser excitation. The photon-exciton detuning was estimated by treating the lowest energy of the measured band structure as the polariton branch which arises due to the coupling between the heavy hole QW exciton (Supplementary Material, Fig. S2) and the photon mode spectrum of a single 3.0 $\mu$m micropillar \cite{Schneider2013}. In doing so, the photon-exciton detuning amounts to -2.3 $\hbar \Omega_R$. The recorded PL spectra are plotted in Figs. 2c-e and depict the dispersion relations along the high symmetry directions K-$\Gamma$-K$^{\prime}$, K$^{\prime}$-M-K and K-K-K, respectively. The optical power density used for the pulsed excitation amounts to $\sim$ 10 Wcm$^{-2}$. Under consideration of nearest-neighbor and second-nearest neighbor coupling, the full dispersion relation of a particle in the S-band of a honeycomb lattice with lattice constant $a$ reads in the TB treatment:
\begin{equation}
 E_{hc}(\textbf{\textit{k}}_{||}) = E_0 \pm t \sqrt{3+f(\textbf{\textit{k}}_{||})} -t' f(\textbf{\textit{k}}_{||}),
\end{equation}
\begin{equation}
f(\textbf{\textit{k}}_{||}) = 2 \cos(\sqrt{3}k_y a) + 4 \cos(\frac{\sqrt{3}}{2} k_y a) \cos(\frac{3}{2} k_x a). 
\end{equation}
All dispersions are successfully described by the TB-fit using one single set of interaction parameters for the nearest-neighbor $t=174$  $\mu$eV and second-nearest neighbor hopping $t'=-26$ $\mu$eV. These values are very similar to the ones found by Jacqmin \textit{et al.} for a fully etched honeycomb lattice ($t=250$ $\mu$eV, $t'=-20$ $\mu$eV) \cite{Jacqmin2014}. In particular, in Fig. 2e the lowest lying band of our dispersion around the energetically degenerate K and K$^{\prime}$ points at the corner of the 1st BZ features the characteristic Dirac cone dispersion of massless particles, stemming from the two-element basis of the lattice. In Fig. 2f, we show a hyperspectral image of the full $(k_{x},k_{y})$-dispersion highlighting the dispersion data points extracted from single measurements (blue) and the corresponding bonding $\pi$ (red) and anti-bonding $\pi^*$ (yellow) orbitals extracted from the TB model. The measured dispersion is in excellent agreement with the model. An energy cut in reciprocal space at the center of the S-band, showing a clear honeycomb geometry with the K and K$^{\prime}$ points, is depicted in Fig. 2g. EL measurements of our polaritonic graphene are depicted in Figs. 2h-j, recorded along the K-$\Gamma$-K$^{\prime}$, K$^{\prime}$-M-K and K-K-K direction, respectively. Here, we apply a voltage yielding an averaged current density of 4.6 A/cm$^2$ flowing through our device. The injected current density in this experiment is well-compatible with strong coupling conditions in our device, as revealed in a control experiment on a standard 20 $\mu$m pillar (Supplementary Material, Fig. S3, which includes Refs. \cite{Schneider2013, Brodbeck2015, Gutbrod1998}). We observe an excellent agreement between the measurements using optical and electrical pumping, and the TB model. More importantly, our electrically injected polariton graphene features well-developed Dirac cones, as well as the flatband dispersion in the P-band.

In order to support the fact that arbitrary lattices can easily be realized, a similar analysis on a square lattice can be found in the Supplementary Material, which includes Refs. \cite{Winkler2016, Schneider2013, Tanese2013,Winkler2016,Sich2018}.

\subsection{Crossing the Threshold to Nonlinear Emission in the Honeycomb Lattice}

\begin{figure}
  \includegraphics[width=0.99\linewidth]{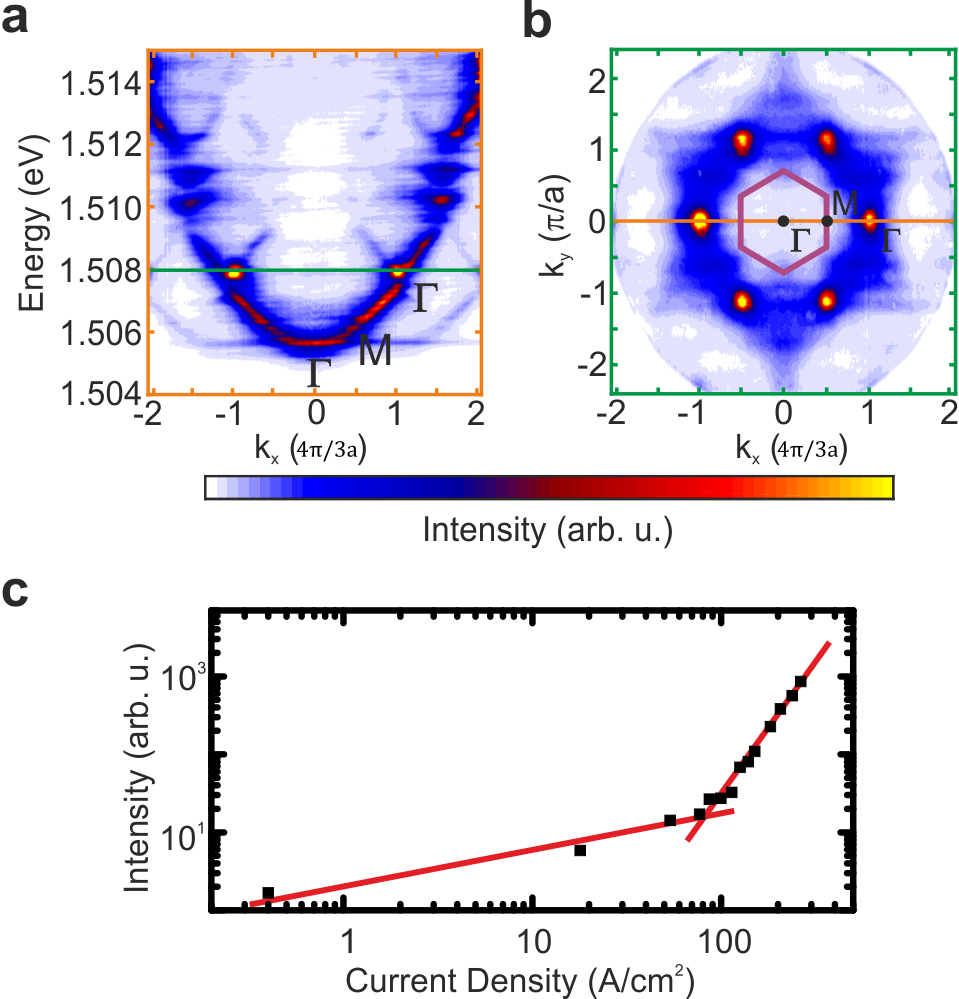}
  \caption{
  (a) Luminescence dispersion of the honeycomb lattice in M-$\Gamma$-M direction under combined electrical and optical excitation. The constant laser power amounts to 90 Wcm$^{-2}$, while the injected current density amounts to 150 A/cm$^{2}$ here. With increasing current density a sharp emission in k-space as well as in energy is seen at the $\Gamma$-point of the 2nd Brillouin zone (BZ) at the bottom of the P-band. (b) Corresponding luminescence dispersion of the honeycomb lattice at a constant energy of 1.508 eV, as indicated by the green line in (a). The 1st BZ and the relevant high-symmetry points are highlighted. (c) Extracted intensity of the relevant mode under combined optical and electrical excitation. A nonlinear increase of the emission intensity is seen around a threshold density of $\sim$ 100 A/cm$^{2}$.}
\end{figure}

A crucial step towards the applicability of our device platform for the use of polaritonic simulation, which facilitates the ultra-fast dynamics of the condensation process, as well as the implementation of topological polariton lasing, involves the demonstration of nonlinear threshold behavior in the input-output response \cite{Amo2016,Berloff2017}. In order to optimally benefit from the narrowband resonance in our cavity device provided by the high photonic quality factor, we extended our study to a far-red-detuned honeycomb device ($\sim$-30 meV). Upon increasing the injected current density, we observe a distinct redshift of the EL spectrum due to ohmic heating below threshold, which prevents the onset of lasing. This can be mitigated by providing a background gain via the incoherent non-resonant Ti:sapphire laser that homogenously illuminates the sample, as specified in the experiment section. The excitation density was chosen to $\sim$ 90 Wcm$^{-2}$, which is safely below the threshold density of $\sim$ 105 Wcm$^{-2}$ at this detuning in the absence of electrical injection. These conditions put the lasing threshold within reach for electrical excitation and upon increasing the injected current density, we observe the onset of well pronounced, monochromatic emission features, which superimpose the luminescence spectrum of the lattice band structure, which is plotted along the M-$\Gamma$-M direction in Fig. 3a. We note that the spots of most intense luminescence correlate with the high-symmetry $\Gamma$ points located in the 2nd BZ of the honeycomb lattice (see Fig 3b), which corresponds to the Bloch-mode within the higher Bloch band (P-band). Apparently, this Bloch-state provides optimal conditions for the maximal gain and results in its strong occupation. We note that polaritonic condensation into higher Bloch states was observed in a 1D chain of trapping potentials \cite{Gao2016}. To analyze the current density dependent emission in more detail, we extracted spectra around $k_x= 4\pi/3a$, and plotted the spectrally integrated intensity as a function of the injected current density in Fig. 3c. The clear threshold characteristic reflects the photonic nonlinearity in our polaritonic graphene, which places our electrically driven device within the zoo of nonlinear photonic on-chip simulators.
\section{\label{sec:level2}CONCLUSION AND OUTLOOK}

\indent In conclusion we have presented the first implementation of an electrically injected, lithographically defined lattice for bosonic quasiparticles, both in the square as well as the honeycomb geometry. The lattice features $\sim$ 250 coupled lattice sites, yielding the prospect for homogeneous driving of extended, arbitrary polariton potential landscapes. The high quality of our fabrication process is reflected in the excellent agreement between photo- and electroluminescence experiments. All acquired data can be reproduced by TB models. We observe well-developed Dirac cones in our electrically driven polariton graphene, besides a flatband dispersion in the P-band. Our device implementation is capable to demonstrate nonlinear behavior under combined electrical and optical excitation, which is a crucial step towards the implementation of bosonic annealing for ultra-fast simulation of the classical $XY$-Hamiltonian \cite{Berloff2017}. We further anticipate that our polariton graphene platform serves well as a platform for the implementation of electrically injected quantum Hall simulators, and related electrically injected topological lasers in the polariton platform \cite{Klembt2018,Ozawa2018}, promising convenient extension to Floquet topological matter \cite{Rechtsman2013}.

\begin{acknowledgments}
We acknowledge support by the ImPACT Program, Japan Science and Technology Agency, the State of Bavaria and by the German Research Foundation (DFG) within project SCHN1376/2-1, SCHN1376/3-1 and KL3124/2-1. S.K. acknowledges the European Commission for the H2020 Marie Sklodowska-Curie Actions (MSCA) fellowship (Topopolis). R.T. is supported by the DFG through SFB 1170 (project B04) and by the European Research Council through ERC-StG-TOPOLECTRICS-Thomale-336012.  S.H. acknowledges support within the EPSRC ”Hybrid Polaritonics” Grant (EP/M025330/1).
\end{acknowledgments}

\end{document}



\title{A platform for electrically pumped \\
polariton simulators and topological lasers\\
Supplementary Material}

\author{Holger Suchomel}
\email{holger.suchomel@physik.uni-wuerzburg.de}
\affiliation{Technische Physik, Wilhelm-Conrad-R\"ontgen-Research Center for Complex
Material Systems, Universit\"at W\"urzburg, Am Hubland, D-97074 W\"urzburg,
Germany}
\author{Sebastian Klembt}
\email{sebastian.klembt@physik.uni-wuerzburg.de\\ H. Suchomel and S. Klembt contributed equally to this work.}
\affiliation{Technische Physik, Wilhelm-Conrad-R\"ontgen-Research Center for Complex
Material Systems, Universit\"at W\"urzburg, Am Hubland, D-97074 W\"urzburg,
Germany}
\author{Tristan H. Harder}
\affiliation{Technische Physik, Wilhelm-Conrad-R\"ontgen-Research Center for Complex
Material Systems, Universit\"at W\"urzburg, Am Hubland, D-97074 W\"urzburg,
Germany}
\author{Martin Klaas}
\affiliation{Technische Physik, Wilhelm-Conrad-R\"ontgen-Research Center for Complex
Material Systems, Universit\"at W\"urzburg, Am Hubland, D-97074 W\"urzburg,
Germany}
\author{Oleg A. Egorov}
\affiliation{Technische Physik, Wilhelm-Conrad-R\"ontgen-Research Center for Complex
Material Systems, Universit\"at W\"urzburg, Am Hubland, D-97074 W\"urzburg,
Germany}
\author{Karol Winkler}
\affiliation{Technische Physik, Wilhelm-Conrad-R\"ontgen-Research Center for Complex
Material Systems, Universit\"at W\"urzburg, Am Hubland, D-97074 W\"urzburg,
Germany}
\author{Monika Emmerling}
\affiliation{Technische Physik, Wilhelm-Conrad-R\"ontgen-Research Center for Complex
Material Systems, Universit\"at W\"urzburg, Am Hubland, D-97074 W\"urzburg,
Germany}
\author{Ronny Thomale}
\affiliation{Institut f\"ur Theoretische Physik, Universit\"at W\"urzburg,
  Am Hubland, D-97074 W\"urzburg, Germany}
\author{Sven H\"ofling}
\affiliation{Technische Physik, Wilhelm-Conrad-R\"ontgen-Research Center for Complex
Material Systems, Universit\"at W\"urzburg, Am Hubland, D-97074 W\"urzburg,
Germany}
\affiliation{SUPA, School of Physics and Astronomy, University of St Andrews, St Andrews
KY16 9SS, United Kingdom}
\author{Christian Schneider}
\affiliation{Technische Physik, Wilhelm-Conrad-R\"ontgen-Research Center for Complex
Material Systems, Universit\"at W\"urzburg, Am Hubland, D-97074 W\"urzburg,
Germany}


\pacs{Valid PACS appear here}
\maketitle


\subsection*{APPENDIX A: SAMPLE DESIGN AND PROCESS}

\indent The pin-doped microcavity structure was grown by molecular beam epitaxy on a GaAs substrate with a (100) orientation and a Si-doping concentration of $\sim$ 2.0$\cdot$10$^{18}$ cm$^{-3}$. Scanning electron microscope (SEM) images of the cleaved edge in Fig. S1a show the full layer structure of the unprocessed sample, in which the n-doped (green), the p-doped (blue) as well as the intrinsic part (red) of the microcavity are highlighted. While silicon was used as a donor impurity in the n-doped part, carbon was used as an acceptor in the p-doped part. The intrinsic $\lambda$-Al$_{0.40}$Ga$_{0.60}$A cavity is surrounded by AlAs/Al$_{0.20}$Ga$_{0.80}$As distributed Bragg reflectors (DBRs). The bottom (top) reflector consists of 31 (27) mirror pairs in which the doping concentration decreases stepwise from 3.0$\cdot$10$^{18}$ cm$^{-3}$ to 0.5$\cdot$10$^{18}$ cm$^{-3}$ towards the cavity as it is shown in the left-hand side graph in Fig. S1a. Besides the highly doped top mirror pairs with a doping concentration of 1$\cdot$10$^{19}$ cm$^{-3}$ to ensure an ohmic contact with the deposited gold in the following process, silicon as well as carbon $\delta$-dopings with an area density of 10$^{12}$ cm$^{-2}$ were introduced in each field minimum between the heterointerfaces in the top and bottom mirror pairs to lower the series resistance of the DBRs. A single stack of four GaAs quantum wells (QWs) with 13 nm width and separated by 4 nm Al$_{0.40}$Ga$_{0.60}$As barriers is placed in the field maximum inside the $\lambda$-Al$_{0.40}$Ga$_{0.60}$As cavity as it is shown in Fig. S1b.\\
\indent The polariton lattices were defined by electron beam lithography and etched into the structure using electron cyclotron resonance-reactive ion etching. The etching was adjusted to stop directly above the cavity region as it is shown in the SEM image of the cleaved edge in Fig. S1c. The cavity region (cyan) is highlighted. In a subsequent step, the structure was planarized with the transparent polymer benzocyclobutene (BCB) which serves as a platform for the following contacting steps. A SEM image of the cleaved edge running through a completed device is depicted in Fig. S1d, showing the planarization and the contact level at the edge of the lattice array. Note that the etching depth outside of the lattice array is deeper than within the holes of the defined lattices, stopping far beyond the cavity region (cyan), due to the larger accessible etching area.

\begin{figure}
  \includegraphics[width=0.95\columnwidth]{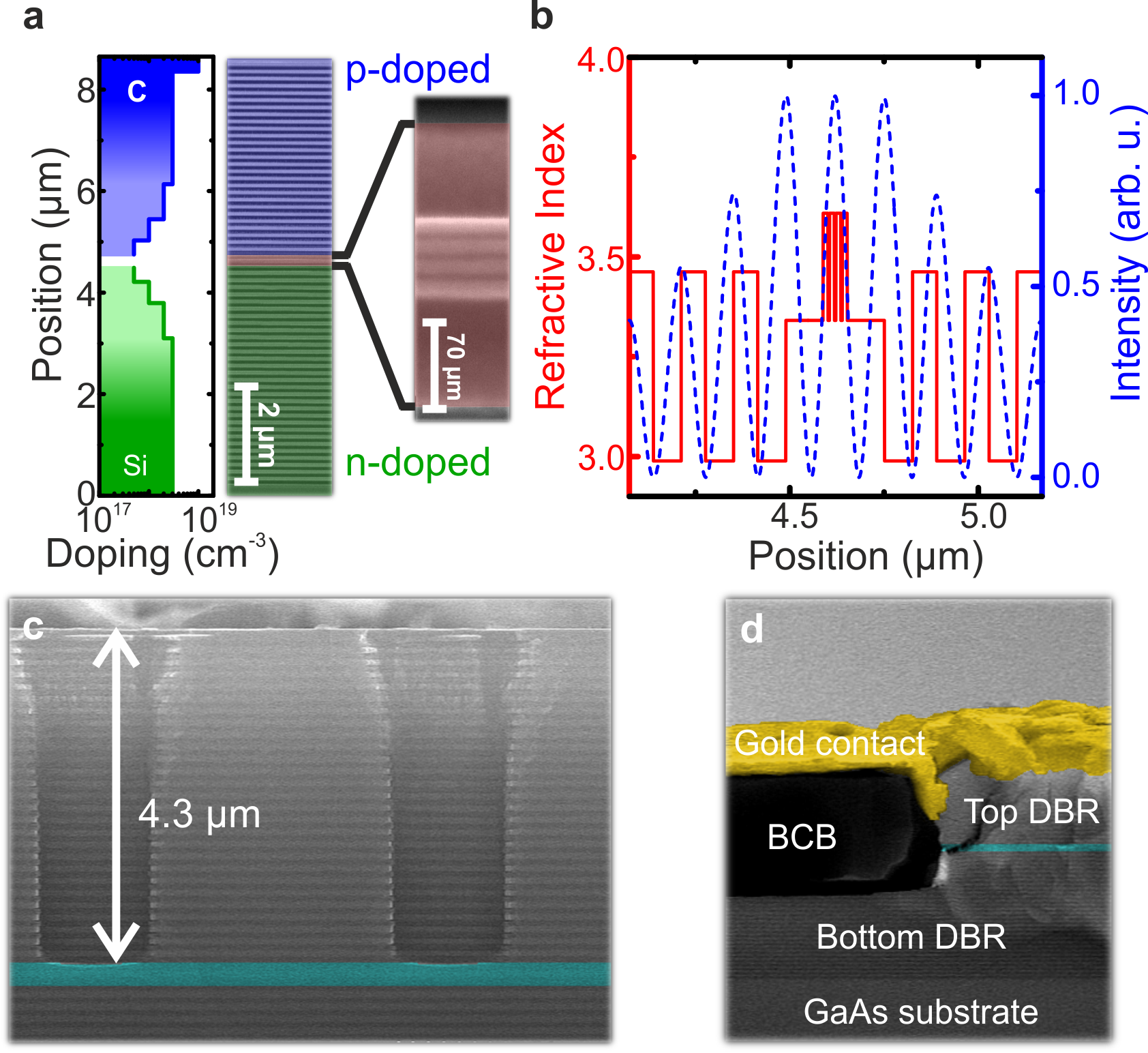}
  \caption{(a) Scanning electron microscope image (SEM) of the cleaved edge showing the pin-doped layer structure of the unprocessed microcavity sample. The blue, red and green area mark the p-doped part, using carbon as a dopant, the intrinsic cavity region as well as the n-doped part, using silicon as a dopant. The graph on the left-hand side shows the exact doping profile, in which the doping concentration decreases step wise towards the cavity. (b) Refractive index distribution (red, solid) and calculated electric field intensity (blue, dashed) of the investigated microcavity structure around the cavity region. A stack of four 13 nm wide GaAs quantum wells is placed in the maximum of the electric field inside the Al$_{0.40}$Ga$_{0.60}$As $\lambda$-cavity. (c) SEM image of the cleaved edge showing the etched honey comb lattice. The etching stops directly above the cavity region (cyan). (d) SEM image of the cleaved edge showing the planarization at the edge of the lattice array.}
\end{figure}

\begin{figure*}
  \includegraphics[width=0.80\linewidth]{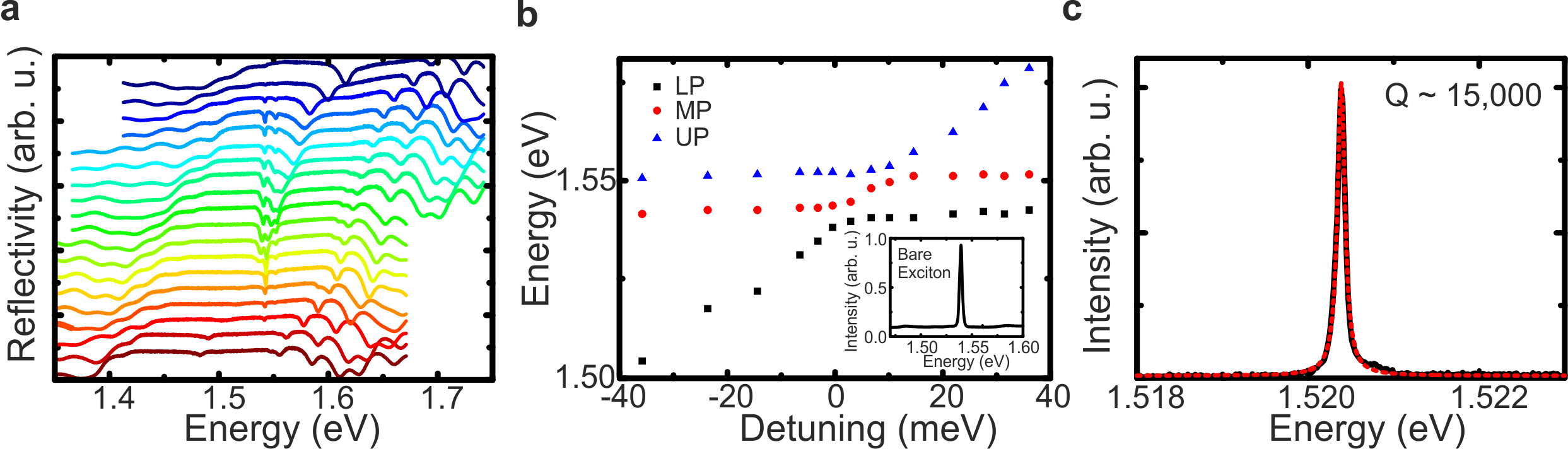}
  \caption{(a) Low temperature ($\sim$ 10 K) white light reflection spectra at different positions on the planar wafer. By utilizing the intrinsic layer thickness gradient along the radial direction of the wafer the cavity mode can be tuned through the exciton energy resonance. To improve the contrast of the absorption resonances, approximately 10 mirror pairs were removed from the top distributed Bragg reflector using reactive ion-etching. (b) Energy position of the reflection dips as a function of heavy-hole exciton-photon detuning. The respective absorption dips can be related to the lower (LP), middle (MP) and upper (UP) polariton branches. The inset shows the photoluminescence spectrum of the bare quantum well-stack. The Rabi-splitting between the LP and the MP branch determines to (5.4 $\pm$ 0.1) meV while the splitting between the MP and the UP branch determines to (4.0 $\pm$ 0.1) meV. The energy position of the photoluminescence peak determines to be 1.539 eV with a linewidth (FWHM) of 2.6 meV. (c) Low temperature ($\sim$ 4 K) photoluminescence spectrum recorded at a far red detuned position on the wafer.}
\end{figure*}

\subsection*{APPENDIX B: BASIC OPTICAL CHARACTERIZATION}

\indent Initially, the planar, unprocessed microcavity sample was characterized in terms of strong coupling, QW photoluminescence (PL) as well as the quality factor (Q-factor) of the microcavity resonator. Concerning strong coupling, a radial stripe from the wafer was investigated via low temperature ($\sim$ 10 K) white-light reflectance measurements. Due to the intrinsic layer thickness gradient along the radial direction of the wafer, the cavity mode can be tuned through the heavy-hole as well as the light-hole exciton energy resonances. Here, the Q-factor was artificially reduced by removing approximately 10 mirror pairs from the top DBR using reactive ion-etching in order to improve the contrast of the absorption resonances. Fig. S2a depicts the recorded reflectance spectra for different positions on the radial stripe. The extracted energy positions of the reflection dips are plotted in Fig. S2b as a function of the heavy-hole exciton-photon detuning. We identify the three individual polariton branches as the lower (LP), middle (MP) as well as the upper (UP) polariton, arising due to the coupling of the cavity mode with the heavy- and light-hole exciton modes, respectively. Both the LP and the MP show a clear anti-crossing with the respective higher polariton branch as the cavity mode is tuned through the respective exciton energy. The minimal energy difference between two polariton branches, that means MP-LP and UP-MP, was identified as the Rabi splitting $\hbar \Omega_R$ and amounts to (5.4 $\pm$ 0.1) meV for the heavy-hole exciton-photon as well as (4.0 $\pm$ 0.1) meV for the light-hole exciton-photon coupling.\\
\indent The inset in Fig S2b shows the PL spectrum of the bare QW-stack at low temperature ($\sim$ 10 K). The whole top DBR of the microcavity structure was removed via reactive ion-etching in order to enter the weak coupling regime. The energy position of the PL peak determines to be 1.539 eV while the linewidth (FWHM) amounts to 2.6 meV. A continuous wave frequency-doubled Nd:YAG laser at an emission wavelength of 532 nm and an excitation power of 5.6 $\mu$W was used. The spot size is determined to be $\sim$ 150 $\mu$m. In Fig. S2c the PL spectrum of the unetched structure is shown from which we extract a Q-factor value of Q $\sim$ 15,000. 

\begin{figure*}
  \includegraphics[width=0.95\linewidth]{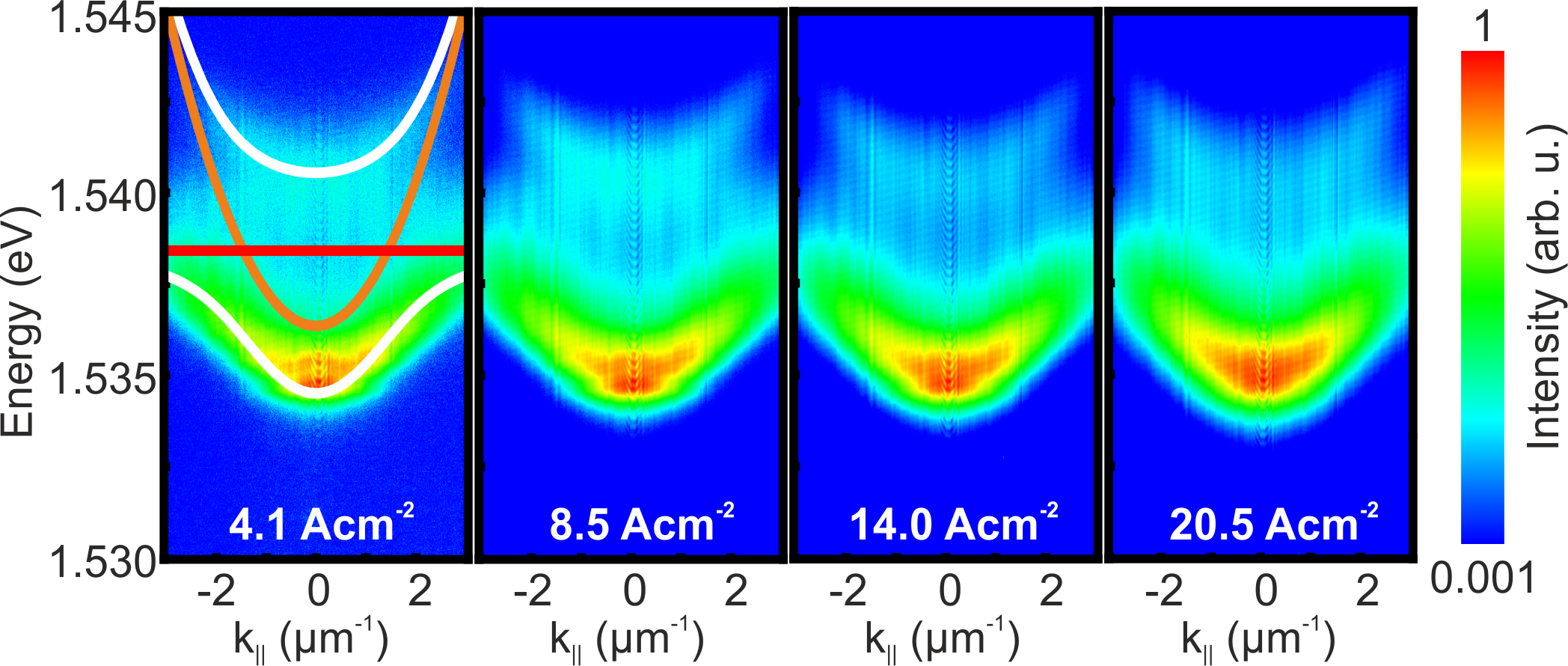}
  \caption{Momentum resolved electroluminescence measurements at different current densities at a single pillar device with a pillar diameter of 20 $\mu$m in a logarithmic color scale and sample temperature of $\sim$ 4 K. At low current densities two anti-crossing branches are visible which we attribute to the lower as well as the upper polariton branch (white lines). Coupling between the photonic ground state of the pillar (orange line) with the heavy-hole exciton (red line) allows fitting the dispersion and gives an exciton-photon detuning of -2.0 meV}
\end{figure*}

\subsection*{APPENDIX C: BASIC ELECTROLUMINESCENCE CHARACTERIZATION}

The grown microcavity structure was further characterized in momentum resolved electroluminescence (EL) measurements at low temperatures ($\sim$ 4 K). For that purpose, single pillar devices with a pillar diameter of 20 $\mu$m and a pn-contact geometry were fabricated, similar to the process described in \cite{Schneider2013, Brodbeck2015}. The momentum resolved spectra at different current densities are shown in Fig. S3. At a current density of 4.1 A/cm$^2$ two anti-crossing dispersions are visible. While the lower one flattens to higher $k_{||}$-vectors the upper one flattens around zero $k_{||}$. We attribute the lower dispersion to the lower (LP) and the upper to the upper (UP) polariton branch. The spectrum was fitted using a standard coupled oscillator model by coupling the photonic ground state of the 20 $\mu$m pillar \cite{Gutbrod1998} with the heavy-hole QW exciton. Assuming a Rabi splitting of 5.4 meV and an exciton energy of 1.538 eV (red line) in accordance with the optical characterization, the cavity mode (orange line) remains the only free fitting parameter. Data fitting yields a heavy-hole exciton-photon detuning of $\delta$ $\sim$ -2.0 meV for the LP and the UP branches (white lines).

\subsection*{APPENDIX D: Square Lattice}

The formation of a band structure by inter-site coupling in lattice configurations with most simplistic geometries is represented by a square lattice. However, it already exhibits interesting phenomena such as localized gap-solitons in a nonlinear polariton regime \cite{Winkler2016}. Here, we study the device depicted in the article in Fig. 1c. The pillar diameter $d=2.5$ $\mu$m and the lattice constant of $a=2.5$ $\mu$m promote a square lattice where the pillars are just touching (Fig. S4a). The corresponding reciprocal lattices is schematically depicted in Fig. S4b and the 1st Brillouin zone (BZ) including the high symmetry points $\Gamma$, X, and M. As a first step, we characterize the optical properties of the lattice by non-resonant optical excitation. The photon-exciton detuning was estimated by treating the lowest energy of the measured band structure as the polariton branch which arises due to the coupling between the heavy hole QW exciton (Fig. S2) and the photon mode spectrum of a single 2.5 $\mu$m micropillar \cite{Schneider2013}. In doing so, the photon-exciton detuning amounts to -2.1 $\hbar \Omega_R$. In Figs. S4c and d, we plot the momentum-resolved PL spectra which were measured along the X-$\Gamma$-X and M-$\Gamma$-M directions in reciprocal space under moderate excitation conditions of $\sim$ 10 Wcm$^{-2}$. Both dispersions are characterized by well-pronounced S- and P-bands arising from the inter-site coupling of neighboring pillars. By scanning the last imaging lens, we can reconstruct the full band structure in the 1st BZ, plotted in Figs. S4e. A cut in energy at the top of the S-band, clearly reflecting the symmetry of the square lattice, is plotted in Fig. S4f. We note that the nearest-neighbor coupling in our structure is sufficient to enable the formation of a fully gapped band structure with a negative effective mass at the BZ-edge (in contrast to e.g. shallow potentials created by metal mask on planar cavities), which is a precondition for further experiments relying on addressing distinct bands, or utilizing the self-focusing nonlinear behavior of quasi-particles with negative effective mass \cite{Tanese2013,Winkler2016,Sich2018}.
More importantly, these primary features of well separated S- and P-bands along the high-symmetry directions X-$\Gamma$-X and M-$\Gamma$-M in reciprocal space are also reflected in the EL spectra plotted in Fig. S4g and h. The effective current density flowing through our device amounts to 4.6 A/cm$^2$.
The success of our technological implementation is further supported by the fact that we can fit both our PL as well as the EL spectra by the identical square lattice tight-binding (TB) model with nearest-neighbor hopping $t$. This dispersion relation derived from solving the linear single-particle Schr\"odinger equation reads:
\begin{equation}
E_{sq}(\textbf{\textit{k}}_{||}) = E_0 + 2t \left\{ \cos(2\pi \, k_x a) +\cos(2\pi \, k_y a) \right\}  
\end{equation}
The fits to the data, based on the identical hopping integral, are shown along with the experimental data in Figs. 4c, d and g, h.

\begin{figure}[h!]
  \includegraphics[width=0.90\linewidth]{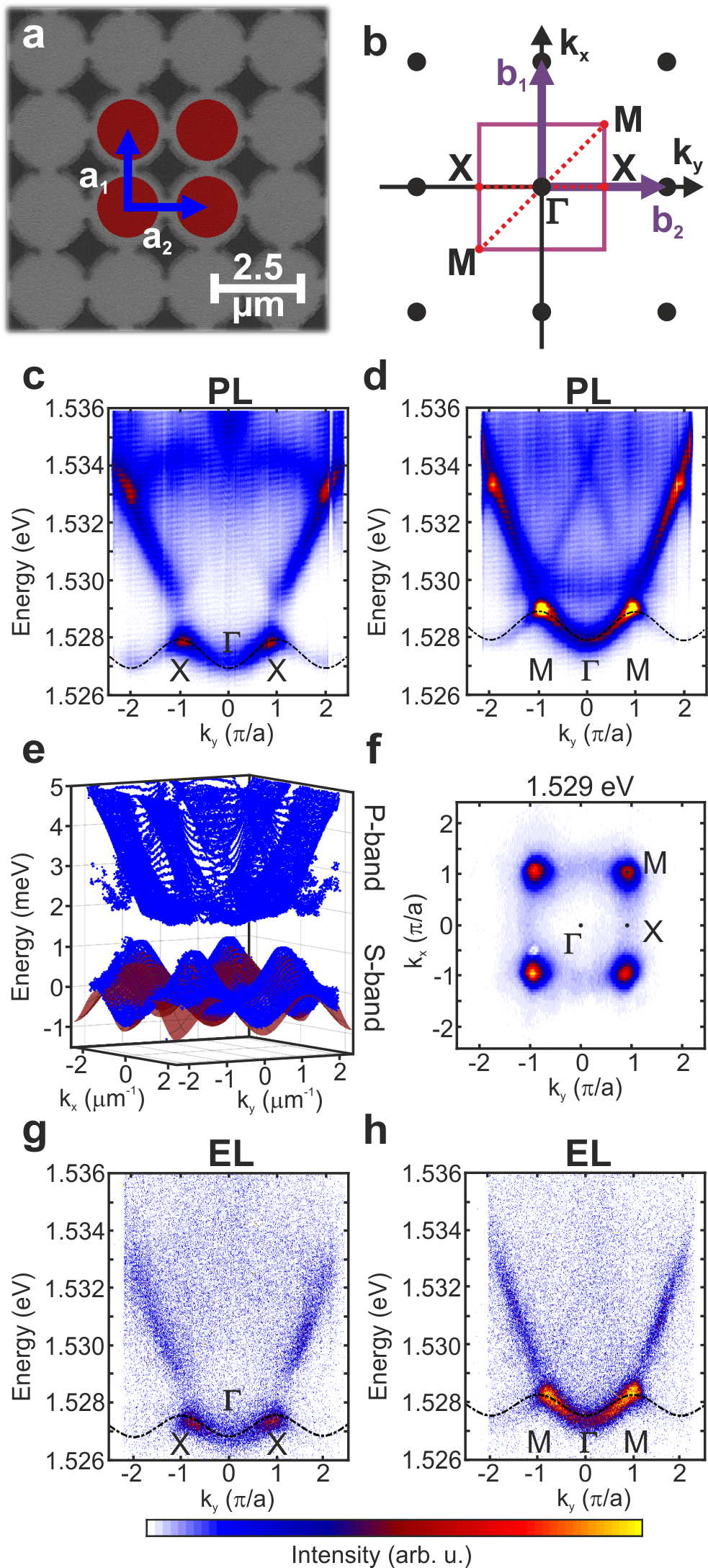}
  \caption{
  (a) Scanning electron microscope image of the investigated square lattice and the corresponding reciprocal space (b). (c,d) PL dispersion in the X-$\Gamma$-X (c) and M-$\Gamma$-M (d) direction of the 1st Brillouin zone (BZ). The dashed line is a tight-binding (TB) calculation reproducing the S-band of the dispersions. The excitation density amounts to $\sim$ 10 Wcm$^{-2}$. (e) Hyperspectral imaging of the 1st BZ with the peak positions fitted to the dispersion spectra (blue) and the TB simulation of the S-band (red). (f) Energy cut in reciprocal space at the top of the S-band, showing a clear square geometry. (g,h) Spectroscopic measurement of the electroluminescence emitted by the very same square lattice in the X-$\Gamma$-X (g) and M-$\Gamma$-M (h) direction of the 1st BZ. The TB calculation uses the same fit parameter as for the photoluminescence measurement.}
 \end{figure}